-----------------------------------------------------------------------

\documentstyle[preprint,aps]{revtex}

\begin{document}
\title{Integral equations of scattering in one dimension}
\author{Vania  E. Barlette$^1$, Marcelo M. Leite$^2$, and Sadhan K.
Adhikari$^3$}

\address{$^1$Centro Universit\'ario Franciscano, Rua dos Andradas 1614,
97010-032 Santa Maria, RS, Brazil\\ $^2$Departamento de F\'isica,
Instituto Tecnol\'ogico de Aeronautica, Centro T\'ecnico Aeroespacial,
12228-900 S\~ao Jos\'e dos Campos, SP, Brazil\\ $^3$Instituto de
F\'{\i}sica Te\'orica, Universidade Estadual Paulista, 01.405-900 S\~ao
Paulo, SP, Brazil\\ } \date{\today} \maketitle

\begin{abstract}
A self-contained discussion of integral equations of  scattering is
presented in the case of centrally-symmetric  potentials in one
dimension, which 
will facilitate the understanding of more complex scattering integral
equations in two and three dimensions. The present discussion illustrates
in a simple fashion the concept of 
 partial-wave decomposition, Green's function,
Lippmann-Schwinger integral equations of scattering for  wave function
and  transition operator, optical theorem and unitarity relation.
We illustrate the present approach with a Dirac delta potential.

\end{abstract}



\newpage

\section{Introduction}

The simple problem of 
one-dimensional quantum scattering continues
as an active line of research \cite{2a}. 
This problem has seen crucial applications 
in studying tunneling phenomena in a finite
superlattice \cite{1}. 
A self-contained discussion of quantum scattering
in one dimension using differential Schr\"odinger equation has appeared
in the literature \cite{1d,1d1,1d2,1d3,1dx}.  Here 
we present a comprehensive  description of the  integral
equations of scattering using the Green's function technique in one
dimension in close analogy with the two- \cite{3} and three-dimensional
cases \cite{4}. Apart from these interests in research, the present study
of 
one-dimensional scattering is also interesting from a pedagogical point of
view. In a one dimensional treatment one does not need special mathematical
functions, like the Bessel's functions, while still retaining sufficient
physical complexity to illustrate many of the physical concepts, such as,
partial-wave decomposition, Green's function,
Lippmann-Schwinger (LS) integral equations of scattering for  wave
function
and  transition operator, optical theorem and unitarity relation,
which
occur in two and three dimensions.

A
self-contained discussion of two-dimensional scattering has appeared in the
literature \cite{3}. There is an intrinsic difference between scattering
in two and three dimensions and that in one dimension. 
In one dimension there are only two
discrete scattering directions: forward and backward along a line. This
requires special techniques in one dimension in distinction with two or
three dimension where an infinity of scattering directions are permitted
characterized by continuous angular variables.
It is because of the above subtlety of one-dimensional scattering problem we
present a complete discussion of the integral-equation formulation in this
case. This special feature of one dimensional scattering leads to two
partial waves. For a centrally-symmetric  potential these two partial
waves are
decoupled, with even and odd parity \cite{1dx}.

Here we present a discussion of the integral equations  of scattering
in one dimension 
in momentum and configuration space for a centrally-symmetric 
potential $V(x)=V(-x)$. 
Form\'anek \cite{1d1} has discussed the 
importance of  this symmetry in obtaining 
mathematical simplification.
Interesting things can happen in the nonsymmetric
case \cite{1d2}. The two partial waves are coupled in the nonsymmetric
case $V(x) \ne V(-x)$ and 
nontrivial modifications of the standard
scattering formulation are needed \cite{1dx}. For noncentral potentials
in
three dimensions (in the presence of tensor potential) the different
partial-wave components also become coupled.  However, in this work we
shall only 
be limited to a discussion of the centrally symmetric case.

More specifically, we present a Green-function description of scattering
in
partial-wave form. 
Also, we derive the
LS equation of
scattering for wave function and a obtain a suitable transition ($t$)
matrix. The on-the-energy shell (on-shell) $t$ matrix element is essentially
the physical scattering amplitude. We also derive the
unitarity relation for the $t$ matrix and show it to be consistent with the
usual form of the optical theorem in three dimensions \cite{4}. 

In Sec. II we
present the LS integral equations. Sec. III deals with the unitary
relations. In Sec. IV we present an illustrative application of the
formalism with the delta function potential. 
A
brief summary is presented in
Sec. V.

\section{Integral equation of scattering}

In one dimension the scattering wave function $\psi _{k}^{(+)}(x)$ at
position $x$ satisfies the Schr\"{o}dinger equation 
\begin{equation}
\left[ -\frac{d^{2}}{dx^{2}}+V(x)\right] \psi
_{k}^{(+)}(x)=k^2\psi _{k}^{(+)}(x)  \label{1}
\end{equation}
where $V(x)$ is a centrally-symmetric finite-range potential in units of
$\hbar^2/(2m)$
satisfying $
V(x)=0,r>R$ and $V(x)=V(-x)$, where $r=|x|$, $m$ the reduced mass, $E\equiv
\hbar ^{2}k^{2}/(2m)$ the energy, and $k$ the wave number. The scattering
wave function is not square integrable and hence does not satisfy the 
usual
normalization condition of a bound-state wave function. However it is
useful in some cases to `normalize' it according to 
\begin{equation}
\int_{-\infty}^{\infty}dx 
\psi _{k}^{(+)*}(x) \psi _{k '}^{(+)}(x)=2\pi  \delta (k-k '),
\end{equation}
where $\delta$ is the Dirac delta function.
 The previous
discussions on scattering in one dimension are based on the solution of
the differential equation (\ref{1}) for the wave function supplemented by
the necessary boundary conditions
\cite{1d,1d1,1d2}. The present discussion will be based on the integral
equations of scattering which we derive below.

In
configuration space the Green's function $G_0^{\pm}\equiv \lim_{\eta \to 
0} (k^2-H_0+i\eta)$, with $H_0$ the kinetic energy operator,  satisfies 
\begin{equation}
\left[ k^2+\frac{d^{2}}{dx^{2}}\right] G_{0}^{(\pm
)}(x,x^{\prime })=\delta (x-x^{\prime })  \label{19}
\end{equation}
together with appropriate boundary conditions. The solution of this equation
is given by \cite{mf}
\begin{equation}
G_{0}^{(\pm )}(x,x^{\prime })=\mp \frac{i}{2k}\exp (\pm
ik|x-x^{\prime }|)  \label{20}
\end{equation}
In analogy with two and three dimensions one can introduce the following
LS  equation 
\begin{equation}
\psi _{k}^{(\pm )}(x)=\phi _{k}(x)+\int_{-\infty}^\infty G_{0}^{(\pm
)}(x,x')V(x')\psi _{k}^{(\pm )}(x') dx'. \label{17}
\end{equation}
In operator form this equation becomes
\begin{equation}\label{17x}
|\psi _{k}^{(\pm )}\rangle =|\phi _{k}\rangle+ G_{0}^{(\pm
)}V|\psi _{k}^{(\pm )}\rangle. 
\end{equation}
The incident plane wave $\phi_{k}(x) \equiv \langle x| k\rangle  
=\exp (ikx)$ satisfies the free
Schr\"{o}dinger equation $H_{0}\phi _{k}=k^2\phi _{k} .$ 
The $\pm $ sign in Eq. (\ref{17}) refers to the outgoing and incoming wave
boundary conditions, respectively.
The physical scattering problem corresponds to the positive sign. 

Equation (\ref{20}) when substituted into  Eq. (\ref{17}) leads to the
following asymptotic form for the wave function  \cite{1dx}
\begin{equation}
\lim_{r\rightarrow \infty }\psi _{k}^{(\pm )}(x)\rightarrow \exp (ikx)+\frac{
i}{k}f_{k}^{(\pm )}(\epsilon )\exp (\pm ikr),  \label{21}
\end{equation}
where
\begin{equation}\label{22a}
f_k^{(\pm)}(\epsilon)
=\mp \frac{1}{2}\int_{-\infty }^{\infty }\exp
(\mp i\epsilon k x^{\prime })V(x^{\prime })\psi _{k}^{(\pm
)}(x^{\prime
})dx^{\prime } . 
\end{equation}
The quantity $f_k^{(\pm)}(\epsilon)$ is termed scattering amplitude.
The parameter $\epsilon$ is $+1$ for $x>0$ and $-1$ for $x<0$ in the
asymptotic region. 
Definition (\ref{22a}) of the scattering amplitude \cite{1d3} is distinct
from
that in Refs. \cite{1d,1d1,1d2}: 
\begin{equation}\label{22ax}
\tilde f_k^{(\pm)}(\epsilon)
= \mp \frac{i}{2k}\int_{-\infty }^{\infty }\exp
(\mp i\epsilon k x^{\prime })V(x^{\prime })\psi _{k}^{(\pm
)}(x^{\prime
})dx^{\prime } . 
\end{equation}
The essential difference between the two is in the phase factor $i$ which
does not influence the cross sections.  However,  the present definition
has  the advantage
of leading to  an optical theorem in close analogy with that in three
dimensions as we shall see in the following. The scattering amplitude of
Refs. \cite{1d,1d1,1d2} leads to
an optical theorem distinct from the one in three dimensions.
The physical outgoing-wave boundary
condition is given by  Eq. (\ref{21}) for $\psi_k^{(+)}(x)$.
There are two discrete scattering directions in one dimension
characterized by
$\epsilon =\pm 1$ in the asymptotic region in contrast to infinite
possibility of 
scattering angles in two and three dimensions. The forward (backward)
direction is given by the $\epsilon = 1$ ($-1$). The discrete
differential cross sections in these two directions are given by 
\begin{equation}
\sigma _{\epsilon }=\frac{1}{k^{2}}|f_{k}^{(+)}(\epsilon )|^{2}  \label{5}
\end{equation}
and the total cross section by \cite{1d,1d1,1d2}
\begin{equation}
\sigma _{\text{tot}}=\sum_{\epsilon }\sigma _{\epsilon }=\frac{1}{k^{2}}
\left[|f_{k}^{(+)}(+)|^{2}+|f_{k}^{(+)}(-)|^{2}\right].  \label{6}
\end{equation}
The discrete sum in  Eq. (\ref{6}) is to be compared with integrals over
continuous angle variables in two and three dimensions. 
The usual reflection ($R$) and transmission ($T$)
probabilities are given by $T=|1+if_k^{(+)}(+)/k|^2$ and 
$R=|if_k^{(+)}(-)/k|^2$.

Here we must comment on the dimension (unit) of cross section and
scattering
amplitudes. Cross section is defined to be the number of events (particles
scattered) per unit time per unit incident flux. The incident flux is
the number of particles incident per unit time per a section of space.
The
section of space has 
units of area and length   in three and  two dimensions
and is 
dimensionless  in one dimension.
The corresponding incident flux is 
measured in units of T$^{-1}$L$^{-2}$, T$^{-1}$L$^{-1}$, and T$^{-1}$ in
these dimensions,
respectively, where T denotes time and L length.  
Cross section is an observable and 
has units of L$^{2}$ and  L in three \cite{4} and two \cite{3} dimensions
and is dimensionless in one
dimension.
The present cross section (\ref{6}) is dimensionless.
Scattering amplitude is not a physical observable and there could be some
flexibility in its definition. The present definition of the scattering
amplitude leads to an optical theorem which is formally similar to
the optical theorem in three dimensions. Other definitions of the
scattering amplitude in terms of
other units are also possible, for example the one given by Eq.
(\ref{22ax}) \cite{1d,1d1}.

Next we introduce the on-shell matrix elements of the transition
operator $t$ via the following equations 
\begin{equation}\label{22}
\langle \epsilon k|t^{(\pm )}(k^2)| k\rangle =- 2 f_{k}^{(\pm
)}(\epsilon )
\end{equation}
\begin{equation}\label{22b}
\langle k^{\prime }|t^{(\pm )}(k^2)|k\rangle 
=\langle \phi _{k^{\prime }}|V|\psi_{k}^{(\pm)}\rangle  
=\int_{-\infty }^{\infty
}\exp
(\mp i k^{\prime }x^{\prime })V(x^{\prime })\psi _{k}^{(\pm
)}(x^{\prime
})dx^{\prime },
\end{equation}
with $k^{\prime}=\epsilon k$ and $\epsilon = \pm 1$, so that
$k^2={k^{\prime}} ^2$; $\epsilon = +1$ ($-1$) corresponds to the forward
(backward) direction. The 
transition operator $t$ is taken to obey  $ t^{(\pm
)}(k^2)|q^{\prime}\rangle
=V|\psi_{q^{\prime}}^{(\pm)}\rangle $ with
${q^{\prime}}^2\ne k^2$.
From Eq. (\ref{17x}) we find that the transition operator $t$ satisfies
\begin{equation}
 t^{(\pm
)}(k^2)= V + V G_0^{(\pm)} t^{(\pm
)}(k^2). \label{xz}
\end{equation} 
The usual spectral representation of the
free Green's function is \cite{mf}  
\begin{equation}
G_{0}^{(\pm )}(x,x')=\frac{1}{2\pi }\int_{-\infty }^{\infty
}dp\frac{\exp (ipx)\exp(-ipx')}{k^2-p^{2}\pm i0 }.  \label{23}
\end{equation}
Using Eqs. (\ref{xz})
and (\ref{23}) one obtains the following explicit LS equation for the
general
off-shell $t$ matrix elements $\langle q|t^{(\pm )}(k^2)|q^{\prime
}\rangle$ ($q^2 \ne k^2 \ne {q^{\prime}}^2$) 
\begin{equation}
\langle q|t^{(\pm )}(k^2)|q^{\prime }\rangle =\langle q|V|q^{\prime
}\rangle
+
\frac{1}{2\pi }\int_{-\infty }^{\infty }dp\langle q|V|p\rangle \frac{1}{
k^{2}-p^{2}\pm i 0 }\langle p|t^{(\pm )}(k^2)|q^{\prime }\rangle. 
\label{24}
\end{equation}
where \begin{equation}
\langle  q|V| q^{\prime }\rangle =\int_{-\infty }^{\infty }\exp
(-i qx)V(x)\exp
(i q^{\prime }x)dx  \label{27}
\end{equation}
Using the symmetry
property
of $V$, e.g. $V(x)=V(-x)$ and Eq. (\ref{27})
it can be shown that 
\begin{equation}
\langle \epsilon q|V|q^{\prime }\rangle =2\int_{0}^{\infty }drV(r)\cos
({\bar{q}^\prime }r+ 
\epsilon {\bar{q}}r),  \label{28}
\end{equation}
where $\bar q=|q|$.
One can conveniently define the momentum-space partial-wave matrix
elements of the potential by 
\begin{equation}
\langle \epsilon q|V|q^{\prime }\rangle =2\sum_{L=0}^{1}\epsilon
^{L}\langle {\bar{q}}
|V_{L}|{\bar{q}^{\prime }}\rangle  \label{26}
\end{equation}
where  $
V_{L}$'s are   the desired partial-wave components. 
Using  Eqs. (\ref{28}) and (\ref{26}) we obtain 
\begin{equation}
\langle {\bar{q}}|V_{0}|{\bar{q}^{\prime }}\rangle
=
\int_{0}^{\infty }dr\cos
({\bar{q}}r)V(r)\cos ({\bar{q}^{\prime }}r)\equiv\frac{1}{2} 
\int_{-\infty}^{\infty }dx\cos
({\bar{q}}x)V(x)\cos ({\bar{q}^{\prime }}x)
  \label{29}
\end{equation}
\begin{equation}
\langle {\bar{q}}|V_{1}|{\bar{q}^{\prime }}\rangle =\int_{0}^{\infty
}dr\sin
({\bar{q}}r)V(r)\sin ({\bar{q}^{\prime }}r).  \label{30}
\end{equation}
Defining the partial-wave elements of the $t$ matrix  as in  
Eq. (\ref{26}) via 
\begin{equation}
\langle \epsilon q|t^{(\pm)}(k^2)|q^{\prime }\rangle
=2\sum_{L=0}^{1}\epsilon
^{L}\langle {\bar{
q}}|t_{L}^{(\pm)}(k^2)|{\bar{q}^{\prime }}\rangle  \label{31}
\end{equation}
one arrives at the following partial-wave projection of the LS equation 
(\ref{24}) 
\begin{equation}
\langle {\bar{q}}|t_{L}^{(\pm )}(k^2)|{\bar{q}^{\prime }}\rangle =\langle
{
\bar{q}}|V_{L}|{\bar{q}^{\prime }\rangle }+\frac{2}{\pi }\int_{0}^{\infty
}d
{\bar{p}}\langle {\bar{q}}|V_{L}|{\bar{p}}\rangle \frac{1}{{{k}}^{2}-{
{p}}^{2}\pm i0 }\langle {\bar{p}}|t_{L}^{(\pm )}(k^2)|{\bar{q}^{\prime
}
}\rangle  \label{32}
\end{equation}
The integration limits in the partial-wave form extend from 0 to $\infty $
as after partial-wave projection we generate the modulus of 
momentum variable.

\section{Unitarity relation}

In order to derive the unitarity relation we write the formal solution of
the LS equation (\ref{32}) as 
\begin{equation}
t_L^{(\pm )}(k^2)=V_L+VG^{(\pm )}(k^2)V_L  \label{33}
\end{equation}
where the full Green's function is defined by 
$G^{(\pm )}(k^2)=\lim_{\eta \rightarrow 0}(k^2-H\pm i\eta )^{-1}  $ with
$H=H_0+V$ the full Hamiltonian.  
Using the complete set of eigenstates of the Hamiltonian $H$
one has the following spectral decomposition of the full Green's
function 
\begin{equation}
G^{(\pm )}(k^2)=\sum_{n}\frac{|\psi _{n}\rangle \langle \psi
_{n}|}{k^2-e_{n}}+
\frac{2}{\pi }\int_{0 }^{\infty }dp\frac{|\psi _{p}^{(+)}\rangle
\langle \psi _{p}^{(+)}|}{k^2-p^{2}\pm i 0 }  \label{35}
\end{equation}
where the summation over $n$ refers to the bound states $\psi
_{n}$ of energy $e_n$
and the integration over $p$ to the scattering states $\psi _{p}^{(+)}$
with the
normalization conditions $\langle \psi _{n}|\psi _{n}\rangle =1$ and $
\langle \psi _{q}^{(+)}|\psi _{p}^{(+)}\rangle =2\pi \delta (p-q).$ From 
Eq. (\ref{35}) one realizes that when this expression is substituted in
Eq. (\ref{33}) 
the integral over the continuum states will contribute a square-root
cut in the complex energy plane along the real positive energy axis, so that
the $t$ matrix will possess this so called unitarity cut. The discontinuity
across this cut is given 
by the following difference 
\begin{eqnarray}
\langle {\bar{q}}|t_{L}^{(+)}(k^2)-t_{L}^{(-)}(k^2)|{\bar{q}}^{\prime
}\rangle
=-4i\int_{0}^{\infty }d{\bar{p}}\delta ( k^{2}-{{p}}^{2})\langle
{\bar{q}}
|t_{L}^{(+)}({{{p}}^{2}})|{\bar{p}}\rangle \langle 
\bar{p}|t_{L}^{(-)}({{p}^{2}})|\bar{q}^{\prime
}\rangle
\end{eqnarray}
or 
\begin{equation}
\Im \langle {\bar{q}}|t_{L}^{(+)}(k^2)|{\bar{q}}^{\prime }\rangle
=-\frac{1}{\bar k}
\langle {\bar{q}}|t_{L}^{(+)}(k^2)|\bar k\rangle \langle \bar 
k|t_{L}^{(-)}(k^2)|\bar{q}
^{\prime }\rangle ,  \label{39}
\end{equation}
where $\Im$ denotes the imaginary part.
When $\bar{q}=\bar{q}^{\prime }=\bar k$, Eq. (\ref{39}) becomes 
\begin{equation}
\Im \langle \bar  k|t_{L}^{(+)}(k^2)|\bar k\rangle =-\frac{1}{\bar 
k}|\langle
\bar k|t_{L}^{(+)}(k^2)|\bar k\rangle |^{2}
\end{equation}
which leads to the following parametrization for the on-shell $t$ matrix
in terms of the phase shift $\delta_L$ 
\begin{equation}\label{xx}
\langle \bar k|t_{L}^{(+)}(k^2)|\bar k\rangle =-\bar k\exp (i\delta
_{L})\sin (\delta _{L}).
\end{equation}
From Eqs.   (\ref{22}), (\ref{31}) and (\ref{xx}) we obtain the following
partial-wave projection of the scattering amplitudes
\cite{1d,1d1,1d2}:
\begin{equation}
f_{k}^{(+)}(\epsilon )=\bar k\sum_{L=0}^{1}\epsilon ^{L}\exp (i\delta
_{L})\sin
(\delta _{L}).  \label{10}
\end{equation}
The two phase shifts for both amplitudes $f_{k}^{(+)}(+)$ and
$f_{k}^{(+)}(-)$
are the same. 
Now one naturally defines the partial-wave amplitudes 
\begin{equation}
f_{L}=\bar k\exp (i\delta _{L})\sin (\delta _{L})  \label{11}
\end{equation}
so that the total cross section of 
Eq. (\ref{6}) becomes \cite{1d,1d1,1d2} 
\begin{equation}
\sigma _{\text{tot}}=2\sum_{L=0}^{1}\sin ^{2}(\delta _{L})  \label{12}
\end{equation}
and satisfies the following optical theorem \cite{1dx}
\begin{equation}
\sigma _{\text{tot}}=\frac{2}{\bar k}\Im f_{k}^{(+)}(+),  \label{13}
\end{equation}
in close analogy with the three-dimensional case, where 
 $f_{k}^{(+)}(+)$ is the forward scattering amplitude. The scattering
amplitude (\ref{22ax}) of \cite{1d,1d1,1d2} leads to the distinct optical
theorem
\cite{1d2}  
\begin{equation}\label{14}
\sigma _{\text{tot}}=-2\Re {\tilde f_{k}^{(+)}(+)},  
\end{equation}
where $\Re$ denotes the real part.

\section{Delta Potential}

Here we present an illustrative application of the formulation using the
delta potential: $V(x)= \lambda \delta(x)$. The configuration space
solution of the problem is well known \cite{gb}. In the following we show
that the integral-equation treatment leads to the same result. The
solutions are analytically known in this simple case.
 From Eqs.
(\ref{29}) and (\ref{30}) we find that 
$\langle \bar q|V_0|\bar q '\rangle = \lambda/2$ and 
$\langle \bar q|V_1|\bar q '\rangle = 0$. Hence we have only one partial
wave ($L=0$) in this case. In this case Eq. (\ref{32}) has the trivial
solution 
\begin{equation}\label{xy}
\langle \bar q|t_0^{(+)}(k^2)|\bar q '\rangle
=\frac {\lambda/2}{1+\frac{\lambda}{\pi}\int _0 ^\infty
\frac{dp}{k^2-p^2+i0} }= \frac{ik\lambda /2}{ik -\lambda/2}.
\end{equation}
In this simple problem the $t$ matrix elements are independent of the
momenta variables. 
From Eqs. (\ref{22}), (\ref{31}) and (\ref{xy}) we find that 
\begin{equation}
\frac{i}{k}f_k ^{(+)}(\epsilon) = \frac{\lambda/2}{ik -\lambda/2},
\end{equation}
for both $\epsilon =+1$ and $ \epsilon =-1$.
Consequently, Eq. (\ref{21}) becomes
\begin{eqnarray}
\lim_{r\rightarrow \infty }\psi _{k}^{(+ )}(x)&\rightarrow& \exp
(ikx)+\frac{
\lambda/2}{ik -\lambda/2}
\exp ( ikr), \label{42}\\
&\rightarrow & \exp
(ikx)+\frac{mv_0}{i\hbar p_0-mv_0}\exp (ikr),\label{43}
\end{eqnarray}
where $\lambda=2mv_0/\hbar^2$ and $\hbar k =p_0$. In Eq. (\ref{43}) we
have
restored the factors of mass $m$ and $\hbar$.  Equation (\ref{43}) is
essentially Eq. (4-95) obtained by configuration space treatment in Ref.
\cite{gb}. Although, in the scattering approach  Eqs. (\ref{42}) and
(\ref{43}) emerge as asymptotic conditions, they are exact results valid
for all $x$ in this simple analytically soluble problem. 
From Eqs. (\ref{xx}) and (\ref{xy}) we find that the phase shift
$\delta_0$ is defined by $\tan \delta_0 = \Im (\delta_0)/ \Re (\delta_0)
= -\lambda/(2k) = -mv_0/(\hbar p_0)$ \cite{gb}. The bound-state energy
corresponds to the pole of the $t$ matrix (\ref{xy}) which happens at
$k^2=-\lambda^2/4$ or at $E=\hbar^2k^2/(2m) = -mv_0^2/(2\hbar^2)$ $-$ a
result well known from configuration-space methods \cite{gb}. As the $t$
matrix and the scattering amplitudes are known, all relevant quantities,
such as cross sections and reflection and transmission probabilities, can
now be  calculated.  The reflection and transmission probabilities are
given by 
$R= |if_k^{(+)}(-)/k|^2 = \lambda^2/(\lambda^2+4k^2)=
mv_0^2
/(2\hbar^2 E+mv_0^2)$,
$T= |1+if_k^{(+)}(+)/k|^2 = 4k^2/(\lambda^2+4k^2)=
E/[E+mv_0^2/(2\hbar^2)]$ \cite{gb},
such that $R+T
=1$,  whereas the cross section is given by $\sigma_{\mbox{tot}}
= |if_k^{(+)}(+)/k|^2+ |if_k^{(+)}(-)/k|^2= 2\lambda^2/(\lambda^2+4k^2)=
2mv_0^2
/(2\hbar^2E+mv_0^2)$.

\section{Summary}

In this paper we have generalized the standard treatments of two- and
three-dimensional scattering to the case of one-dimensional scattering. We
have introduced the concept of scattering amplitude, partial wave, phase
shift, unitarity, Lippmann-Schwinger scattering equation for the wave
function and the $t$ matrix  in close analogy
with three-dimensional scattering. In this case there are two partial
waves: $L=0 $ and 1 in contrast to two and three
dimensions where there is an  infinite number of partial waves. In one
dimension the
algebra is
much simpler and mostly analytic. 
These make the integral equation formulation
of scattering in one dimension along with its 
partial-wave projection 
easily tractable.   Consequently, difficult concepts of scattering, such
as optical theorem and unitarity,  are easily understandable. Hence the
present formulation will be helpful for  teaching formal theory of
scattering in two and three dimensions.

The work is supported in part by the Conselho Nacional de Desenvolvimento -
Cient\'{i}fico e Tecnol\'{o}gico, Funda\c{c}\~{a}o de Amparo \`{a} Pesquisa
do Estado de S\~{a}o Paulo, and Finan\-ciadora de Estu\-dos e Projetos of
Brazil.

\end{document}